\documentclass[aps,prl,nofootinbib,twocolumn,superscriptaddress,floatfix]{revtex4}
\usepackage{graphicx}
\usepackage{amsmath}

\begin{document}

\title{Direct detection and CMB constraints on light DM scenario of  top quark asymmetry and dijet excess at Tevatron}

\author{Andi Hektor}
\affiliation{NICPB, R\"avala 10, 10143 Tallinn, Estonia}
\author{Gert H\"utsi}
\affiliation{Tartu Observatory, 61602 T\~oravere, Estonia}

\author{Mario Kadastik}
\affiliation{NICPB, R\"avala 10, 10143 Tallinn, Estonia}
\author{Kristjan Kannike}
\affiliation{Scuola Normale Superiore and INFN, Piazza dei Cavalieri 7, 56126 Pisa, Italy}
\affiliation{NICPB, R\"avala 10, 10143 Tallinn, Estonia}

\author{Martti Raidal}
\affiliation{NICPB, R\"avala 10, 10143 Tallinn, Estonia}

\author{David M. Straub}
\affiliation{Scuola Normale Superiore and INFN, Piazza dei Cavalieri 7, 56126 Pisa, Italy}

\date{\today}

\begin{abstract}
We study in detail the model by Isidori and Kamenik that is claimed to
explain the top quark forward-backward asymmetry at Tevatron, provide GeV-scale dark matter (DM), and possibly improve the agreement between data and theory
in Tevatron $W + jj$ events. We compute the DM thermal relic
density, the spin-independent DM-nucleon scattering cross section, and the cosmic microwave
background constraints on both Dirac and Majorana neutralino DM in the parameter space that explains
the top asymmetry. A stable light neutralino is not allowed unless the local DM
density is 3-4 times smaller than expected, in which case Dirac
DM with mass around 3 GeV may be possible, to be tested by the Planck
mission. The model predicts a too broad excess in the dijet distribution and a strong modification of the
missing $E_{T}$ distribution in $W + jj$ events.
\end{abstract}


\maketitle

\section{Introduction}

There are two questions in modern particle physics that have got more attention  than any other. 
The first one is the origin of electroweak symmetry breaking. 
The second one is the nature of the cold dark matter (DM) of the Universe. 
The recent CDF measurement of an anomalously large $t \bar t$ asymmetry~\cite{Aaltonen:2011kc,CDF2} 
and  the observation of a peak in the dijet invariant mass distribution in association with a $W$ boson~\cite{Aaltonen:2011mk}  have been most unexpected and seemingly unrelated to these questions.
Therefore the proposed explanations to those results often involve rather exotic physics meant to explain  the CDF results alone.

One very interesting exception in the long list of proposed models is the one by Isidori and Kamenik \cite{Isidori:2011dp}.
Although the model was proposed to explain the anomalous forward-backward (FB) $t\bar t$ asymmetry, it involves ``usual"
particles, stops and neutralinos, that may originate from  supersymmetric models like the next-to-minimal supersymmetric standard model (NMSSM). Therefore the model provides a DM 
candidate with mass of order 2-3~GeV. (The CoGent experiment recently announced possible evidence for light DM \cite{Aalseth:2011wp}, but their region of spin-independent direct detection cross section is several magnitudes lower than in the model \cite{Isidori:2011dp}.) Such a light DM is very difficult to test even in the most sensitive DM direct detection experiments such as
XENON100 and, at first sight, seems to be experimentally allowed~\cite{Aprile:2011hi} (see \cite{Farina:2011bh} for an up-to-date analysis of the XENON results). 
In addition, the authors claim that the model gives a  non-resonant
contribution to the $W+jj$ channel that may improve the agreement between data and theoretical expectations.
It is intriguing that with a rather minimalistic set of new particles with well-defined couplings and masses one can simultaneously explain
the $t\bar t$ FB asymmetry, the DM of the Universe as well as address the observed $W+jj$ excess.

The aim of this paper is to  study the model \cite{Isidori:2011dp} in detail and to work out the related phenomenology.
We incorporate two versions of the model into the MicrOMEGAs~\cite{Belanger:2006is,Belanger:2010gh}
 package allowing the neutralino to be either a Dirac or a Majorana fermion.
While the collider phenomenology  is not sensitive to the nature of a DM fermion, cosmological observables like the DM abundance and 
DM annihilation cross section at $T=0$ depend crucially on it. We compute the DM abundance and spin-independent
direct detection cross section per nucleon with MicrOMEGAs by 
scanning over the particle masses and couplings that explain the CDF $t \bar t$ FB asymmetry. We find that, indeed, for the model parameters allowed by the asymmetry, the observed DM thermal relic abundance~\cite{Larson:2010gs}
 can be generated and DM as light  as 2-3~GeV is possible if its couplings to quarks approach non-perturbative 
values. Such a light DM is constrained by two observables. First, the observed lack of distortions in the Cosmic Microwave Background (CMB)
spectrum due to DM annihilations at redshifts $z\sim 500$-$1000$ puts a stringent constraint on GeV-scale DM  independently of 
cosmological uncertainties such as the DM  halo profiles, densities and distribution~\cite{Hutsi:2011vx}. 
Following Ref.~\cite{Hutsi:2011vx} we compute the constraints on
$\langle\sigma_{\text{A}} v\rangle(T=0)$ for the annihilation channel $\bar \chi\chi\to \bar u u$ and show that present CMB measurements 
disfavor a light Dirac neutralino as the DM. The Majorana neutralino annihilation cross section at $T=0$ is proportional to the $u$-quark mass~\cite{Beltran:2008xg}
and is not constrained. Second, the spin independent DM-nucleon scattering cross section is so large that 
CRESST-I~\cite{Angloher:2002in}, TEXONO \cite{Lin:2007ka} and XENON100~\cite{Aprile:2011hi} experiments together  exclude
  the light DM possibility in this model entirely. 
Also, $\chi$ cannot be a subdominant component of DM because this would require increasing its 
annihilation cross section and, consequently,  the direct detection cross section by the same amount. We conclude that the model
proposed by Isidori and Kamenik can be a viable model for explaining the $t\bar t$ FB asymmetry but the neutralino $\chi$ cannot be a stable particle and cannot contribute to the DM abundance. 

We note that a re-analysis of the experiments \cite{Savage:2008er}
presents conservative bounds on detector sensitivities. In this case, if, in addition, the local DM density in the location of the Earth is a few times smaller than the present estimates, a small Dirac neutralino parameter space opens up for $m_\chi\sim 3$~GeV. The exact measurement of the CMB by the Planck mission will be the definitive test for this possibility, except when the neutralino is a strongly subdominant component (below 30\%) of DM.

In addition, we study whether the model can reproduce the dijet excess observed by CDF.
We find that the model indeed yields an excess of the $W+jj$ signature but the invariant mass distribution of the dijet is so broad that it cannot explain the CDF observation of a peak, and the escaping neutralinos lead to a strong modification of the missing $E_T$ distribution.

\section{The model}
In addition to standard model (SM) particles, the model comprises the color triplet scalar $\tilde{t}$ and neutral fermion $\chi^0$ that are both $SU(2)$ singlets.
In the case $\chi^0$ is a Dirac fermion, the Lagrangian is
\begin{equation}
\begin{split}
\mathcal{L_{\text{Dirac}}} &= \mathcal{L}_{\text{SM}} + (D_{\mu} \tilde{t})^{\dagger} (D^{\mu} \tilde{t}) 
  - m_{\tilde{t}}^2 \tilde{t}^{\dagger} \tilde{t} + \bar{\chi}^0 (i \gamma_\mu D^\mu) \chi^0 \\
  &-m_\chi \bar{\chi}^{0} \chi^0 - \sum_{q = u,c,t} \left( \tilde{Y}_q \bar{q}_R \tilde{t} \chi^0 + \text{H.c.} \right),
\end{split}
  \label{eq:L:Dirac}
\end{equation}
while in the case of Majorana $\chi^0$, the Lagrangian reads
\begin{equation}
\begin{split}
\mathcal{L_{\text{Maj}}} &= \mathcal{L}_{\text{SM}} + (D_{\mu} \tilde{t})^{\dagger} (D^{\mu} \tilde{t}) 
  - m_{\tilde{t}}^2 \tilde{t}^{\dagger} \tilde{t} + \frac{1}{2} \bar{\chi}^0 (i \gamma_\mu D^\mu) \chi^0 \\
  &- \frac{1}{2} m_\chi \bar{\chi^c}^{0} \chi^0 - \sum_{q = u,c,t} \left( \tilde{Y}_q \bar{q}_R \tilde{t} \chi^0 + \text{H.c.} \right).
\end{split}
  \label{eq:L:Maj}
\end{equation}

In numerical computations, the used ranges for parameters were $1.2 \leq \tilde{Y}_u \leq \pi$, $4 \leq \tilde{Y}_{t} \leq 2 \pi$, $0.1~\text{GeV} \leq m_\chi \leq 20~\text{GeV}$ and $194~\text{GeV} \leq m_{\tilde{t}} \leq 215~\text{GeV}$. To avoid bounds from flavor physics, we take $\tilde{Y}_c = 0$. The $(m_{\tilde{t}}, \tilde{Y}_u)$ pairs used were selected from the region compatible with the $\bar{t}t$-asymmetry at $2\sigma$ level in Fig.~1 of \cite{Isidori:2011dp}, extrapolated up to the value $\tilde{Y}_u = \pi$. We stress that our conclusions do not depend on the details of extrapolation. A variation of $\tilde{Y}_{t}$ is added for completeness; as long as $\tilde{Y}_{t} \gg \tilde{Y}_{u}$, correct $t\bar{t}$ asymmetry is achieved \cite{Isidori:2011dp}, but the exact value of $\tilde{Y}_{t}$ 
only influences the $W+jj$ and missing $E_{T}$ distribution as shown below. Of course, different values of $\tilde{Y}_{t}$ up to $4 \pi$ imply different scales for new physics.

\section{Cosmological observables }

\begin{figure}[hb]
\begin{center}
\includegraphics[width=0.5\textwidth]{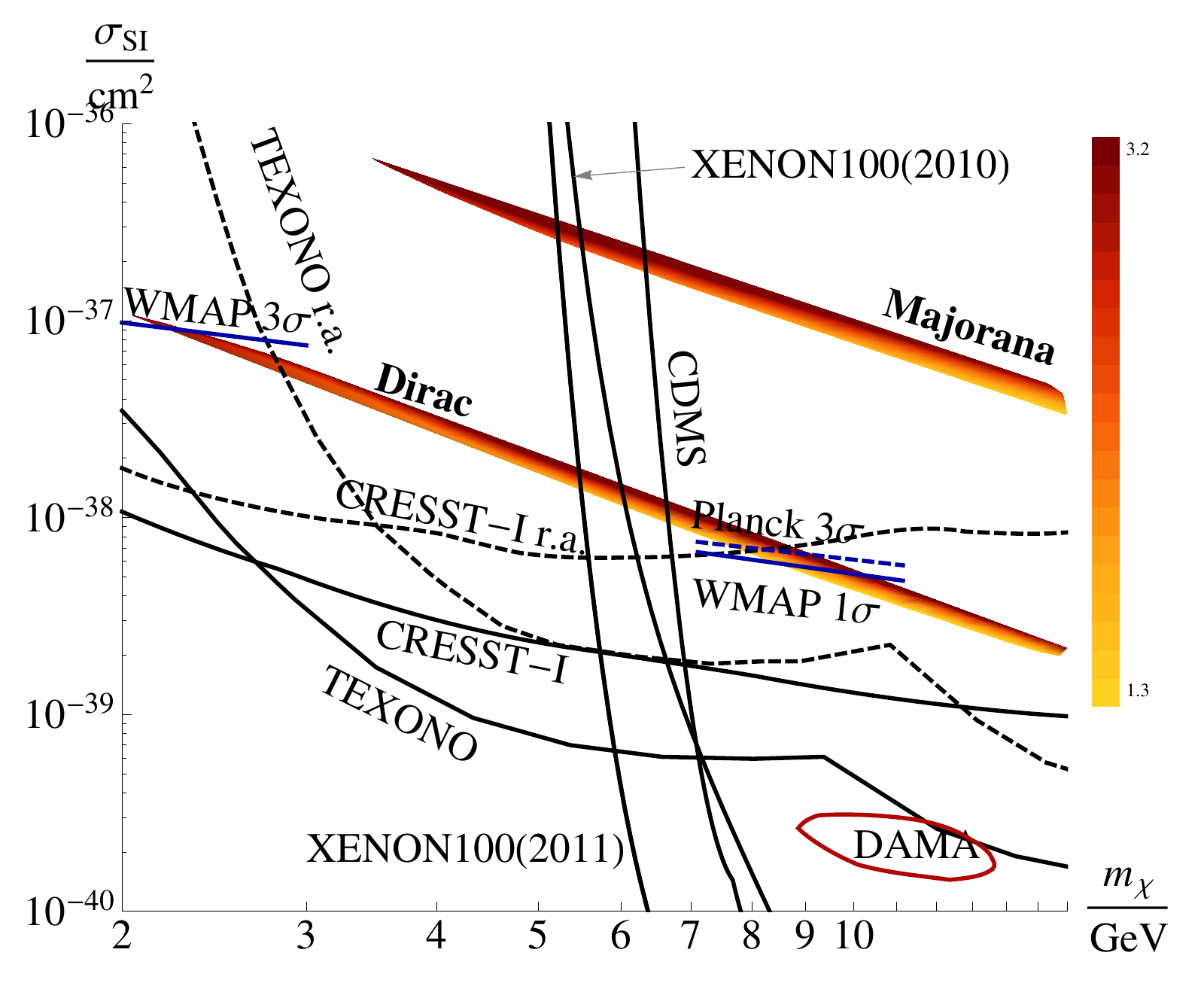}
\caption{Spin-independent nucleon-DM scattering cross section $\sigma_{\text{SI}}$ (scaled by $\Omega_\chi/\Omega_{\text{DM}}$) vs DM mass $m_{\chi}$ for Dirac and Majorana DM.
The strength of $\tilde{Y}_u$ coupling is represented by the color (greyscale) code. 
XENON100~\cite{Aprile:2011hi}, CRESST-I~\cite{Angloher:2002in} and TEXONO \cite{Lin:2007ka} constraints as well as  DAMA~\cite{Bernabei:2008yi} 
best fit region are presented. The reanalyzed (r.a.) bounds for CRESST-I and TEXONO \cite{Savage:2008er} are shown with dashed lines. The CMB constraints (blue lines) apply only to the Dirac $\chi$ and are given for the case when it is the dominant component of DM.  }
\label{fig:sigma:si:mchi}
\end{center}
\end{figure}

The most interesting result of the paper  \cite{Isidori:2011dp} is that the model simultaneously explains  the
measured top quark FB asymmetry at Tevatron and provides the correct thermal relic abundance of DM.  
In order to compute the observables related to DM numerically as precisely as possible we incorporated 
both Lagrangians, Eqs. \eqref{eq:L:Dirac} and \eqref{eq:L:Maj}, into MicrOMEGAs code by using the FeynRules package \cite{Christensen:2008py} for Mathematica. While for collider observables at Tevatron
there should not be major differences between these two cases, the cosmological observables depend crucially on the nature of 
fermionic DM. Because a Majorana particle can have only effective four-fermion axial-vector interaction~\cite{Goodman:2010yf}, 
both the behavior and the magnitude of the DM annihilation cross section will be different in the two cases, as well as the DM scattering cross section with matter.

We scan over the model parameters as explained above and compute the resulting DM relic abundance and 
spin-independent DM-nucleon scattering cross section. All the generated points can explain the CDF $t\bar t$ FB asymmetry.
Requiring $\Omega_{\rm DM} h^{2}<0.1288,$ which is smaller than the upper limit of the WMAP7 $3\sigma$ range, we plot in Fig.~\ref{fig:sigma:si:mchi} 
the computed spin-independent nucleon-DM scattering cross section $\sigma_{\text{SI}}$  as a function of the 
 DM mass $m_{\chi}$ for both Dirac and Majorana DM. The magnitude of the coupling $\tilde{Y}_u$ is represented with the color code.
 Notice that we do allow the model to provide just a subdominant component of the DM of the Universe. As detector sensitivity lines are given for the central value of DM abundance, the cross section of points is scaled down by the fraction of $\chi$ DM in all DM, namely, by $\Omega_\chi/\Omega_{\text{DM}}$.
However, because the annihilation cross section of $\bar\chi\chi\to \bar uu$ and the direct detection cross section of $\chi u\to \chi u$
both scale with  $\tilde{Y}_u$, the prediction is rather sharp.
In the same figure we also give the present DM direct detection constraints.
While XENON100 excludes  DM with mass above 5-6 GeV, CRESST-I and TEXONO exclude all the allowed parameter space both for Dirac as well as for
Majorana DM. Although the sensitivities of CRESST-I and TEXONO are not comparable with the XENON100 one, their bounds extend down to 1 GeV because as a target
CRESST-I used Al and TEXONO used Ge that are much lighter than Xe.

Those model predictions can be readily understood with the help of analytic expressions for annihilation cross section 
obtained in \cite{Beltran:2008xg}. 
In order to allow the DM mass to be in GeV region, the cross section of the annihilation process
$\bar\chi\chi\to\bar uu$ and, consequently, the coupling $\tilde{Y}_u$, must be large. This explains also why the direct detection cross sections are large. 
While the Dirac fermion cross section
contain both $s$- and $p$-wave contributions, the Majorana DM annihilation is purely $p$-wave and, in the limit of vanishing $u$-quark mass, vanishes for
vanishing velocities. Therefore, to obtain a large enough cross section for the Majorana neutralino the coupling $\tilde{Y}_u$ must 
be very large. Consequently the spin-independent direct detection cross section for the Majorana case is larger than in the Dirac case.

Because the interpretation of  results of the direct detection experiments depends on the local DM density in the neighborhood of the Solar System, 
one can make an attempt to save the model by appealing to the conservative reanalyzed sensitivity bounds for CRESST-I and TEXONO given by \cite{Savage:2008er} and shown in Fig.~\ref{fig:sigma:si:mchi} with dashed lines, and reducing the DM density a few times. Indeed, cosmological uncertainties 
in determining the DM halo of our galaxy may be larger than expected and the model prediction for Dirac neutralino is not too
far from the CRESST-I bound. However, it was shown in  \cite{Hutsi:2011vx}  that the present WMAP7 CMB measurements imply strong constraints on the 
annihilation cross section of the GeV-scale DM.  Those constraints are free of uncertainties of DM halo shapes and distributions but apply 
for the cross sections at temperatures $T\sim 0.$ In our case the CMB measurements constrain the Dirac DM case only because the
Majorana DM annihilation cross section vanishes for the reionization temperature $T = 10^{-5}~\mathrm{GeV} \approx 0$.

\begin{figure}[t]
\begin{center}
\includegraphics[width=0.45\textwidth]{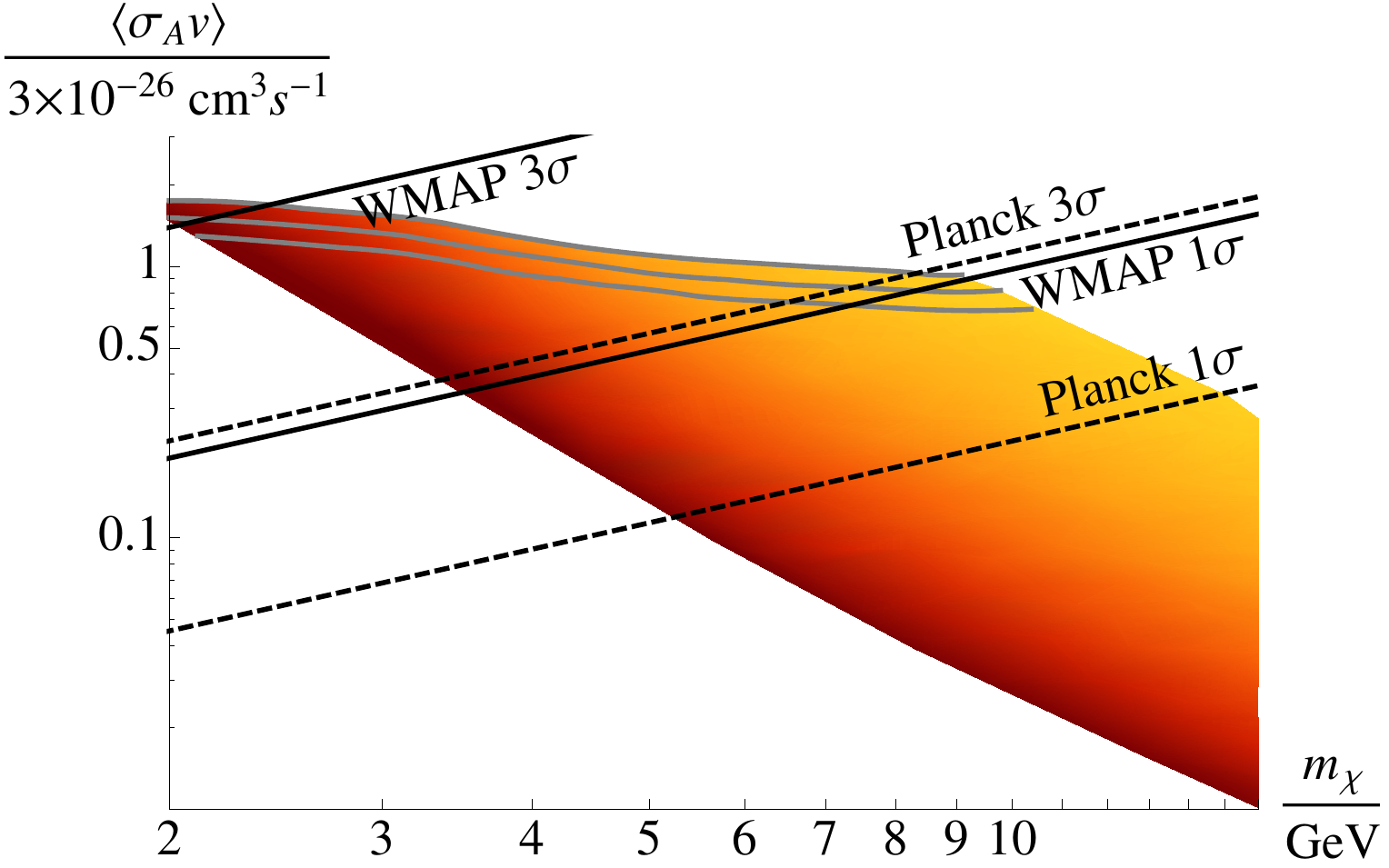}
\caption{CMB constraints on $\langle \sigma_{\text{A}} v \rangle$ (scaled by $\Omega_\chi^{2}/\Omega_{\text{DM}}^{2}$) of Dirac DM at $T=0$  as a function of DM mass $m_{\chi}$ for the annihilation channel 
$\bar\chi\chi \to \bar u u $.  The coloured (shaded) area is the model prediction with the color code for $\tilde{Y}_u$ as in Fig.~\ref{fig:sigma:si:mchi}. Grey lines show the central value of $\Omega_{\text{DM}}$  and the WMAP7 $3 \sigma$ bounds.}
\label{fig:sigma:v:mstop}
\end{center}
\end{figure}

We repeated the analyses of  Ref.~\cite{Hutsi:2011vx} for the channel $\bar\chi\chi\to \bar u u$. The constraints in the $(m_\chi, \langle\sigma_{\text{A}} v\rangle (T=0))$  
plane are presented in Fig.~\ref{fig:sigma:v:mstop} together with the area of model predictions using the same color code 
 as in Fig.~\ref{fig:sigma:si:mchi}. The lines in the figure present WMAP7 
$1~\sigma$ and $3~\sigma$ bounds and the predicted sensitivity of the Planck mission \cite{Zaldarriaga:2008ap} for the central value of $\Omega_{\text{DM}}$. Because the intensity of annihilation is proportional to the square of density, we scale $\langle\sigma_{\text{A}} v\rangle$ by the factor of $\Omega_\chi^{2}/\Omega_{\text{DM}}^{2}$. The corresponding CMB constraints are also
shown in Fig.~\ref{fig:sigma:si:mchi} for $\chi$ as the dominant component of DM. Consequently, if the detector bounds could be reduced at least by 3-4 times by assuming that the local DM density is smaller than expected (in effect, detection sensitivity lines shift up by the same factor), at $2~\sigma$ level a very small
parameter space at $m_\chi = 3.3$-$3.6$~GeV for Dirac DM opens up. However, the variability of the local density of DM is unknown experimentally and very poorly estimated by computer simulations at the spatial scales of the Solar system \cite{Springel:2008cc, Gao:2010tn}. Thus we are unable to give a quantitative estimation for the probability of having local density of DM that is a few times smaller, but such a reduction is highly unlikely (e.g. \cite{Maciejewski:2010gz}). This parameter region will be definitively tested by the Planck mission, except when neutralino DM constitutes less than 30\% of all DM.

To allow Majorana DM to be viable, the local DM density must be reduced by 2 orders of magnitude, which is not realistic.

\section{\boldmath $W$ + Dijet distribution at Tevatron}

The CDF collaboration recently published data on the invariant mass distribution of jet pairs produced in association with a $W,$ finding an excess of events around 150~GeV after background subtraction \cite{Aaltonen:2011mk}. The D0 collaboration, on the other hand, has seen no such signal \cite{Abazov:2011af}.

As noted in \cite{Isidori:2011dp}, the model contributes to this signature via stop pair production,
\begin{equation}
u\bar u \;\; \to \;\; \tilde t (\to \chi \bar u) \; \tilde t(\to \bar\chi t(\to W^+b)) \,.
\end{equation}
We simulate this channel as well as the Standard Model diboson contribution
for two benchmark points with $\tilde Y_t=3$ and 4, respectively, fixing in both cases $\tilde Y_u = 1.5$ and $m_{\tilde t} = 205~\text{GeV}$ to reproduce the $t\bar t$ FB asymmetry and $m_\chi = 2~\text{GeV}$ (the results are weakly dependent on the precise value of $m_\chi$).
We use MadGraph~5 \cite{Alwall:2007st} for parton-level event generation, Pythia~6 \cite{Sjostrand:2006za} for hadronization and PGS~4 width the CDF parameter set for an estimate of detector effects. We apply the kinematical cuts described in \cite{Aaltonen:2011mk} at the detector level.

\begin{figure}[t]
\begin{center}
\includegraphics[width=0.5\textwidth]{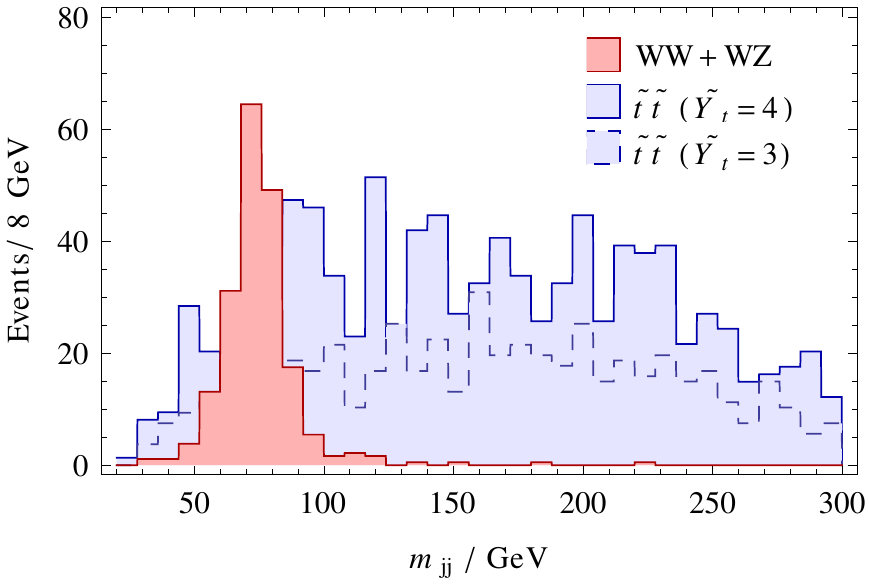}
\caption{Dijet invariant mass $m_{jj}$ distribution in the two benchmark scenarios (blue, light grey) and the Standard Model diboson contribution (red, dark grey).}
\label{fig:Mjj}
\end{center}
\end{figure}

The distribution of the dijet mass $m_{jj}$ for the Standard Model and the two benchmark scenarios are
shown in Fig.~\ref{fig:Mjj}.
The model indeed leads to an excess in the same region as seen in the CDF analysis.
However, while we do not directly compare the simulation to the data, in view of our imperfect detector simulation, lack of next-to-leading order and jet energy scale corrections, it is obvious that the distribution is much broader than the CDF bump and that the excess thus extends to much lower and higher values of $m_{jj}$.

\begin{figure}[t]
\begin{center}
\includegraphics[width=0.5\textwidth]{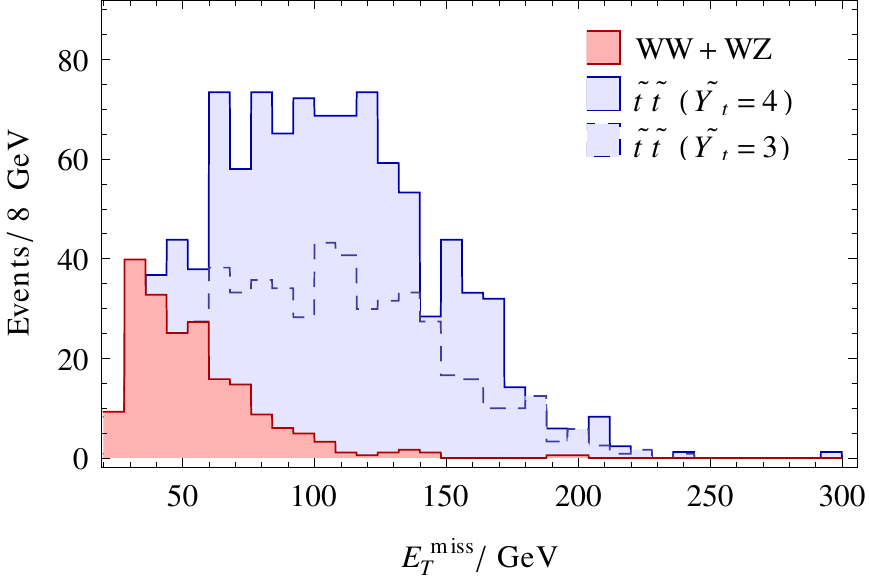}
\caption{Missing $E_T$ distribution in the two benchmark scenarios (blue, light grey) and the Standard Model diboson contribution (red, dark grey).}
\label{fig:MET}
\end{center}
\end{figure}

In addition, the two neutralinos present in the final state lead to a significant increase in the missing energy per event. We show the missing $E_T$ distribution, after applying the same cuts as for Fig.~\ref{fig:Mjj}, for the diboson contribution and the two benchmark points in Fig.~\ref{fig:MET}. While the corresponding background-subtracted plot from the experiment is not public to our knowledge, it should clearly show this huge excess, to be tested at the LHC as well (the CMS detector resolution for missing $E_{T}$ is around 10 GeV \cite{etmiss}).

We conclude that, while the model might slightly reduce the tension between theory and data for the $m_{jj}$ distribution, the strong modification of the missing $E_T$ distribution excludes this possibility.

\section{Conclusions}

We have performed a detailed study of the Isidori and Kamenik model \cite{Isidori:2011dp}. We find that the DM direct detection experiments and the CMB measurement by WMAP7
exclude the possibility that the Dirac or Majorana neutralino could contribute to the DM relic abundance. 
If the local DM density in the Solar System is reduced 3-4 times (a rather unlikely possibility), a small parameter region for $m_\chi = 3.3$-$3.6$~GeV may open for DM 
at $2~\sigma$ level. This possibility will be tested by the Planck mission, except when neutralino DM makes up less than 30\% of all DM.
Alternatively the model must be modified so that the neutralino $\chi$ decays before or after nucleosynthesis. 
In this case the model still provides  a viable scenario to explain the $t\bar t$ FB asymmetry but not  DM.
The model cannot explain the CDF dijet anomaly due to a too broad dijet invariant mass distribution and significant additional missing transverse energy in $W+jj$ events.

\paragraph{Acknowledgements.}
We thank Riccardo Barbieri for several discussions and Alexander Pukhov for providing us with a new version of MicrOMEGAs package.
This work was supported by the ESF grants  8090, 8499, 8943, MTT8, MJD52, MJD140 and by SF0690030s09 project, and by the EU ITN "Unification in the LHC Era", contract PITN-GA-2009-237920 (UNILHC).

\end{document}